\documentclass{JHEP3}

  \usepackage{latexsym,bm,amsmath,amssymb,amsfonts}
  \usepackage{epsfig,graphics,graphicx}
  \usepackage{slashed}
  \usepackage{cite}
  \usepackage[latin1]{inputenc}

  \long\def\comment#1{ }

  \newcommand{\beq}{\begin{eqnarray}}
  \newcommand{\eeq}{\end{eqnarray}}
  
 \def\simge{\mathrel{%
   \rlap{\raise 0.511ex \hbox{$>$}}{\lower 0.511ex \hbox{$\sim$}}}}
\def\simle{\mathrel{
   \rlap{\raise 0.511ex \hbox{$<$}}{\lower 0.511ex \hbox{$\sim$}}}}

\vspace{1.5cm}

\preprint{
}

\title{\rm \LARGE Polarized DIS in ${\mathcal N}=4$ SYM: Where is spin at strong coupling?}

\author{Yoshitaka Hatta and Takahiro Ueda\\Graduate School of Pure and Applied Sciences, University
of Tsukuba, Tsukuba, Ibaraki 305-8571, Japan\\
E-mail: \email
{hatta@het.ph.tsukuba.ac.jp}, \email{tueda@het.ph.tsukuba.ac.jp
 }}
\author{Bo-Wen Xiao\\ Nuclear Science Division, Lawrence Berkeley National Laboratory, Berkeley, CA 94720, USA\\
 E-mail: \email {BXiao@lbl.gov}}

\abstract{Using the AdS/CFT correspondence, we calculate the polarized structure functions in strongly coupled ${\mathcal N}=4$ supersymmetric Yang--Mills theory deformed in the infrared. We find that the flavor singlet contribution to the $g_1$ structure function is vanishingly small, while the flavor non--singlet contribution shows the Regge behavior at small--$x$ with an intercept slightly less than 1. We explicitly check that the latter satisfies the moment sum rule. We discuss the `spin crisis' problem and  suggest that at strong coupling the spin of a hadron entirely comes from the orbital angular momentum.
}

\begin{document}

\section{Introduction}

The question of how the spin of the proton  is distributed among quarks and gluons has been one of the central themes in QCD. It all started in the late 1980s when  the European Muon Collaboration (EMC) announced data for polarized deep inelastic scattering (DIS) which apparently suggested that a disturbingly small fraction of the proton's total spin is accounted for by the helicity of quarks  \cite{Ashman:1987hv}.
 After more than twenty years of vigorous experimental and theoretical efforts since then, the problem of the missing spin---or the `spin crisis'---has now become much less mysterious than it was initially thought. [See \cite{Kuhn:2008sy} for the latest review of the subject.] Nevertheless, there still remains a great deal of controversy  regarding the nature of the decomposition of the total spin
\beq \frac{1}{2}=\frac{1}{2}\Delta \Sigma + \Delta G + L_z\,, \label{frac}
\eeq
where $\Delta \Sigma$ and $\Delta G$ are contributions from the helicity of quarks and gluons, respectively, and $L_z$ is their orbital angular momentum. Among the three terms in (\ref{frac}), $\Delta \Sigma$ is relatively well--constrained by a wealth of DIS data supplemented by the NLO global QCD analysis.
 The value often quoted in the literature is  $\Delta \Sigma \sim 0.25$.
 This is  larger than the original EMC result, but is still significantly smaller than predictions based on the naive quark model $1\ge \Delta \Sigma \gtrsim 0.6$.
  On the other hand, extraction of $\Delta G$ is more difficult as it is not directly measurable in DIS. While there have been several other experiments which are better suited to this purpose, at present the data  are not yet sufficient (especially in the low--$x$ region) for an accurate determination of $\Delta G$. However, one may notice that some of the recent analyses \cite{Alexakhin:2006vx,Leader:2006xc,Abelev:2007vt,Adare:2008px,deFlorian:2008mr,Hirai:2008aj,deFlorian:2009vb} prefer a small value of $\Delta G$ consistent with zero, albeit with large uncertainties of order 0.5.

The present status of the sum rule (\ref{frac}) just described is a persistent reminder of fundamental theoretical questions in spin physics: (i) While the evolution with $Q^2$ of the helicity distributions $\Delta \Sigma(Q^2)$ and $\Delta G(Q^2)$ are calculable in perturbation theory at large $Q^2 \gg \Lambda^2_{QCD}\,$, they are essentially nonperturbative quantities. Can one evaluate them in the strong coupling regime where the perturbative approach breaks down? (ii) Why is $\Delta \Sigma$ `unnaturally' small, and what carries the rest of the total spin?  If $\Delta G$ is small as suggested by some recent works, it may well be that the orbital angular momentum $L_z$ is the dominant component.
Existing models \cite{Brodsky:1988ip,Schreiber:1988uw,Myhrer:1988ap,Wakamatsu:1999nj}  that can give rise to a large value of $L_z$ include the nonperturbative dynamics of QCD in one way or another, so again the key to understand this question resides in the strong coupling sector.
(iii) How do the polarized parton densities and structure functions behave at small values of the Bjorken $x$ variable? In addition to its phenomenological impact in fitting the data at low--$x$, the question is of great conceptual interest because one expects, as in the unpolarized case, an interplay between  `hard' and `soft' Regge exchanges \cite{Bass:2006dq}.

 In this paper, we try to address these questions using the AdS/CFT correspondence \cite{Aharony:1999ti} which has emerged as a powerful tool to analyze  the strong coupling regime of non--Abelian gauge theories. Specifically, we shall study polarized DIS in  ${\mathcal N}=4$ supersymmetric Yang--Mills  (SYM) theory at large 't Hooft coupling $\lambda \gg 1$ and compute structure functions in the small--$x$ region.
 In the context of gauge/string duality, DIS was first formulated in \cite{Polchinski:2002jw} for the unpolarized case (see, also \cite{Hatta:2007he,Hatta:2007cs,BallonBayona:2007qr,BallonBayona:2007rs,Cornalba:2008sp,Pire:2008zf}), and  recently generalized to the polarized case in \cite{Gao:2009ze}. The analysis of \cite{Gao:2009ze} is  focused on the large--$x$ region which corresponds to the supergravity approximation on the string theory side. However, the internal degrees of freedom (`partons') of hadrons are visible only at extremely small values of $x$ of order  $\sim e^{-\sqrt{\lambda}}$  which in fact is  the Regge regime of the theory \cite{Polchinski:2002jw}. Thus, the very existence of spin physics at strong coupling is intrinsically tied to high energy Regge scattering, and this is what we are going to explore.

In ${\mathcal N}=4$ SYM, there are four `flavors' of Weyl fermions associated with the SU(4) ${\mathcal R}$--symmetry.  We shall find that the flavor singlet contribution to the $g_1$ structure function is simply \emph{vanishing} for all values of $Q^2>\Lambda^2$ where $\Lambda$ is an infrared scale (analog of $\Lambda_{QCD}$) necessary to break conformal symmetry
\beq
\Delta \Sigma=0\,. \label{sk}
\eeq
 Interestingly, the same conclusion has been reached in a Skyrme model calculation \cite{Brodsky:1988ip}. [Note that both approaches use the large--$N_c$ approximation.]  As a matter of fact, in our case (\ref{sk})  follows rather trivially from general arguments. Instead, our main focus is the flavor non--singlet contribution which  shows the expected power--law increase in $1/x$ as $x\to 0$.
  In the context of gauge/string duality, this of course is attributed to the usual Regge behavior of the string S--matrix.

  In Section~2, we review the standard perturbative approach to polarized DIS in QCD. In Section~3 we consider the same process in ${\mathcal N}=4$ SYM and identify the dominant $t$--channel process that survives at strong coupling. The actual calculation of the structure function is performed in Section~4 where we present an explicit expression of the $g_1$ function (\ref{find}) which exactly satisfies the moment sum rule. Possible implications of our results for QCD spin physics will be discussed at the end.

\section{Review of polarized DIS in QCD}
In this section we briefly review the basics of the operator product expansion (OPE) approach to polarized structure functions in QCD.\footnote{ In accordance with the majority of the QCD literature, we shall employ the `mostly minus' metric signature $(+---)$ for the Minkowski space and also in the gravity calculations in later sections.}

We start with the forward Compton amplitude off a polarized hadron (usually a proton) target
\beq T^{\mu\nu}=\frac{i}{2\pi} \int d^4 y e^{iqy} \langle PS|T\{J^\mu(y)J^\nu(0)\}|PS\rangle = T_{sym}^{\mu\nu}+iT_{asym}^{\mu\nu}\,,
\label{1}\eeq
where $P^\mu$ and $S^\mu$ are the momentum and spin vectors, respectively, and $J^\mu=\sum_f e_f\bar{q}_f\gamma^\mu q_f$ is the electromagnetic current of quarks. The subscript $sym/asym$ refers to the symmetric/antisymmetric part under the exchange of Lorentz indices  $\mu\leftrightarrow \nu$.
On the other hand, the structure functions are the components of the following hadronic tensor
\beq
W^{\mu\nu}=\frac{1}{2\pi} \int d^4y e^{iqy}\langle PS|J^\mu(y)J^\nu(0)|PS\rangle &=&\frac{1}{2\pi}\int d^4y e^{iqy}\langle PS|[J^\mu(y),J^\nu(0)]|PS\rangle  \nonumber \\
&=&W^{\mu\nu}_{sym} +iW_{asym}^{\mu\nu}\,.
\eeq
After some manipulations one can show that $W^{\mu\nu}_{sym}$ and $W^{\mu\nu}_{asym}$ are real, and that
\beq
2\,{\rm Im}\, T_{sym}^{\mu\nu}&=&W_{sym}^{\mu\nu}\,, \\
2\,{\rm Im} \,T_{asym}^{\mu\nu}&=&W_{asym}^{\mu\nu}\,, \label{22}
\eeq
 which represents the optical theorem.
The polarized structure functions $g_1$ and $g_2$ are contained in $W_{asym}^{\mu\nu}$. From (\ref{1}) and (\ref{22}), one has, for the antisymmetric part, ($\epsilon_{0123}=-\epsilon^{0123}=+1$)
\beq
&& {\rm Im}\, \frac{1}{2\pi}\int d^4y e^{iqy}\langle PS|T\{J^\mu(y)J^\nu(0)\}| PS \rangle\Big\arrowvert_{asym} = \frac{1}{2}W_{asym}^{\mu\nu} \nonumber \\ &&=\epsilon^{\mu\nu\alpha\beta}q_\alpha \left(\frac{S_\beta}{P\cdot q}(g_1(x,Q^2)+g_2(x,Q^2))-\frac{q\cdot S P_\beta}{(P\cdot q)^2}g_2(x,Q^2) \right)\,, \label{start}
\eeq
where $S^2=-M^2$ with $M$ being the hadron mass.

 The usual strategy to evaluate the structure function is to perform an OPE for the product $JJ$
 \beq
 &&\int d^4y\, e^{iqy}\langle PS|T\{J^\mu(y)J^\nu(0)\}| PS \rangle\Big\arrowvert_{asym} \nonumber \\
&& \qquad =\frac{2}{Q^2}\epsilon^{\mu\nu\alpha\beta}q_\alpha \sum_{j=1}^{odd}\frac{2q^{\mu_1}\cdots 2q^{\mu_{j-1}}}{Q^{2(j-1)}} \sum_f e_f^2\langle PS|\Theta^f_{\{\beta,\mu_1,\cdots \mu_{j-1}\}}+\Theta^f_{[\beta \{\mu_1]\mu_2,\cdots \mu_{j-1}\}}|PS\rangle \nonumber \\
&& \qquad = \epsilon^{\mu\nu\alpha\beta}\frac{q_\alpha}{P\cdot q} \sum_j^{odd}\left(\frac{a_j+(j-1)d_j}{j}\frac{S_\beta}{x^{j}}
 +\frac{(j-1)(a_j-d_j)}{j}\frac{ S\cdot q}{P\cdot q}\frac{P_\beta}{x^j}\right)\,, \label{inf}
 \eeq
  where we defined the operators
  \beq
  \Theta^f_{\{\beta,\mu_1,\cdots \mu_{j-1}\}} &=&\frac{1}{j}\bar{q}_f\gamma^5\gamma_{\{\beta}iD_{\mu_1}\cdots iD_{\mu_{j-1}\}}q_f\,, \label{111}
\\ \Theta^f_{[\beta \{\mu_1]\mu_2,\cdots \mu_{j-1}\}}&=&
  \frac{1}{j}\sum_i^{j-1}\left(\bar{q}_f\gamma^5\gamma_\beta iD_{\{\mu_1}\cdots iD_{\mu_{j-1}\}}q_f -\bar{q}_f\gamma_5\gamma_{\mu_i}iD_{\{\mu_1}\cdots iD_\beta \cdots iD_{\mu_{j-1}\}}\right)\,, \label{222} \nonumber \\
  \eeq
   and $a_j$, $d_j$ are the coefficients in their matrix elements. [The brackets $\{\cdots\}$ ($[\cdots]$) denote the (anti--)symmetrization and trace--subtraction of Lorentz indices.]

Although each term in (\ref{inf}) is real, one finds an imaginary part after summing over $j$ and  analytically continuing in $x$ to  the physical region $0\le x\le 1$. The result is
\beq
g_1(x)&=&\frac{1}{2}\frac{P^+}{4\pi S^+}\sum_f e^2_f \Bigl(\int dy^- e^{ixP^+y^-}\langle PS|\bar{q}_f(0)\gamma_5\gamma^+W[0,y^-]q_f(y^-)|PS\rangle  \nonumber \\ && \qquad \qquad  + \int dy^- e^{-ixP^+y^-}\langle PS| \bar{q}_f(0)\gamma_5 \gamma^+W[0,y^-]q_f(y^-)|PS\rangle \Bigr)
\nonumber \\ &\equiv& \frac{1}{2}\sum_f e_f^2 (\Delta q_f(x)+\Delta \bar{q}_f(x))\,,
\eeq
  where $W[0,y^-]$ is the Wilson line operator that assures gauge invariance and in the second equality we have introduced the polarized quark and antiquark distributions.

Taking $x$--moments of the structure functions, one obtains  sum rules
\beq
\int_0^1 dx\, x^{j-1}g_1(x,Q^2)&=&\frac{a_j}{4}\,,   \label{g1} \\
\int_0^1 dx\, x^{j-1}g_2(x,Q^2)&=&-\frac{(j-1)}{4j}(a_j-d_j)\,,
\eeq
  which can be inverted to give a well--known relation  \cite{Wandzura:1977qf}
  \beq
  g_2(x,Q^2)=-g_1(x,Q^2) + \int_x^1 \frac{dz}{z}g_1(z,Q^2) + \bar{g}_2(x,Q^2)\,. \label{ww}
  \eeq
  The last term $\bar{g}_2$ comes from the operator (\ref{222}) and represents the twist--three contribution.  (\ref{ww}) shows that the twist--two part of $g_2$ is related to $g_1$.
 For $j=1$, the sum rule (\ref{g1}) measures the axial charge $a$ of the target. One finds
\beq \int_0^1 dx\, g_1(x,Q^2) =\frac{1}{4S^+}\sum_f e_f^2 \langle PS|\bar{q}_f \gamma_5 \gamma^+ q_f|PS\rangle =\frac{1}{9}a^{(0)}+\frac{1}{12}a^{(3)}+\frac{1}{36}a^{(8)}\,, \label{term}
\eeq
where $a^{(0)}$ is the flavor singlet axial charge and $a^{(3,8)}$ are the octet (non--singlet) charges. [To leading order, the coefficient functions are set to 1.] In massless QCD, the flavor octet axial currents are exactly conserved, so $a^{(3,8)}$ are independent of $Q^2$. However, the flavor singlet axial current is not conserved due to anomaly. This induces a weak $Q^2$--dependence in $a^{(0)}$ in the NLO approximation \cite{Kodaira:1979pa} where one also finds ${\mathcal O}(\alpha_s)$ corrections in the coefficient functions \cite{Kodaira:1979ib}.
 One can eliminate $a^{(0,8)}$ by taking a difference between the proton and neutron structure functions
  \beq
 \int_0^1 dx \left( g_1^p(x,Q^2)-g_1^n(x,Q^2)\right)=\frac{1}{12}(a^{(3)}_p-a^{(3)}_n)=\frac{g_A}{6}\,,
  \label{bjo}
 \eeq
  where $g_A \approx 1.25$ is the isovector axial charge of the nucleon.
 (\ref{bjo}) is known as the Bjorken sum rule \cite{Bjorken:1966jh} and has been very well tested  in experiment.

 The singlet axial charge $a^{(0)}$ is nothing but the quark's helicity contribution to the total spin
 \beq a^{(0)}= \sum_f \bigl( \Delta q_f + \Delta \bar{q}_f \bigr)
=\sum_f \bigl( q^+_f - q^-_f+\bar{q}^+_f -\bar{q}_f^- \bigr) = \Delta \Sigma\,, \label{heli}
\eeq
where $\Delta q_f=\int_0^1 dx \Delta q_f(x)$ and the superscript $+/-$ means that the helicity is parallel/antiparallel to the proton spin.
As already mentioned in Introduction, roughly the spin crisis is tantamount to the unnaturally small value of $a^{(0)}=\Delta \Sigma$ as compared to unity and also to the non--singlet axial charges.

\section{Polarized DIS in ${\mathcal N}=4$ SYM}
From this section on we discuss polarized DIS in ${\mathcal N}=4$ supersymmetric Yang--Mills theory at large values of the 't Hooft coupling $\lambda  \gg 1$. Due to the AdS/CFT correspondence, the problem reduces to the scattering of closed strings in the product space of five--dimensional anti--de Sitter space AdS$_5$ and the 5--sphere
 \beq
 ds^2=\frac{2dy^+dy^- - dy_1^2-dy_2^2-dz^2}{z^2}-d\Omega_5^2\,, \label{metri}
 \eeq
  where we set the common radius $R$ of AdS$_5$ and S$^5$ equal  to unity. In this unit, the string tension becomes  $\alpha'=\frac{1}{\sqrt{\lambda}}$.
 Since ${\mathcal N}=4$ SYM is not confining, one has to break conformal symmetry to generate hadrons to be used as a target in DIS. On the string theory side, this can be done by deforming the ten--dimensional metric such that it approaches (\ref{metri}) near the Minkowski boundary $z\to 0$, but differs from it far away from the boundary, around  $z\sim 1/\Lambda$. The characteristic scale $\Lambda$ then sets the mass scale of hadrons.  In this way one can ensure that the theory is identical to  ${\mathcal N}=4$ SYM in the ultraviolet (UV). This is essential to our discussion  since we shall heavily rely on the OPE  which of course is a property in the UV.
On the other hand, there is an important caveat regarding  possible forms of the deformation in the infrared region $z\sim 1/\Lambda$. We shall comment on this as we go along.

 The basic strategy to calculate the structure functions has been laid out in \cite{Polchinski:2002jw} for the unpolarized case and in \cite{Gao:2009ze} for the polarized case. Here we briefly recapitulate the essential features at strong coupling and then discuss the extension to the polarized case.  In the ${\mathcal N}=4$ theory, the analog of the electromagnetic current would be a 15--component exactly conserved current $J_a^\mu$ ($a=1,\cdots, 15$) associated with the global SU(4) ${\mathcal R}$--symmetry. [Its explicit form is given in Appendix A.] One then introduces a photon field by gauging a U(1) subgroup of SU(4) which, for definiteness, is taken to correspond to  the generator $t^{a=3}={\rm diag}(1/2,-1/2,0,0)$. As for the target, one employs a spin--$\frac{1}{2}$ hadron which has mass $M\sim \Lambda$ and charge under SU(4).   Similarly to (\ref{start}), one can then define structure functions
\beq
&& {\rm Im}\, \frac{1}{2\pi}\int d^4y e^{iqy}\langle PS|T\{J^\mu_3(y)J^\nu_3(0)\}| PS \rangle\Big\arrowvert_{asym}  \nonumber \\ &&=\epsilon^{\mu\nu\alpha\beta}q_\alpha \left(\frac{S_\beta}{P\cdot q}(g_1(x,Q^2)+g_2(x,Q^2))-\frac{q\cdot S P_\beta}{(P\cdot q)^2}g_2(x,Q^2) +\frac{P_\beta}{2P\cdot q}F_3(x,Q^2) \right)\,, \label{tens}
\eeq
 where the parity--violating structure function $F_3$ is expected to arise since the theory is chiral.

For large values of the Bjorken--$x$, the physical picture is completely different from QCD. Instead of the twist--two operators (\ref{111}) which acquire a large anomalous dimension of order $\lambda^{1/4}$, the OPE is dominated by the double trace operators which create and annihilate the entire hadron. These operators are nominally of higher twist, but since their dimension is protected, at moderately large $Q^2$ they give a larger contribution to the structure function than the twist--two operators. Thus in this regime it is not possible to probe the internal structure of hadrons, but rather the hadron appears as a pointlike particle (see, however, \cite{Pire:2008zf}).

When $x$ is  exponentially small, $x\sim e^{-\sqrt{\lambda}}$, the scattering starts to look more like the situation in QCD at least from the viewpoint of the OPE. In the unpolarized case, the scattering is dominated by the graviton exchange which is dual to the energy momentum tensor. This is a twist--two operator with spin $j= 2$, and its dimension is protected. Because of the curvature of the AdS space,  the relevant value of $j$  is slightly shifted away from 2. However,  the anomalous dimension remains of order unity in the vicinity of $j=2$, and thus the twist--two operators give the leading contribution.

Similarly, in the polarized case   one expects that some  protected operator in the $J_3J_3$ OPE gives the more important contribution than the double trace operators considered in \cite{Gao:2009ze}. We observe that this operator is the ${\mathcal R}$--current itself. Our argument is the following: There exists an exact (i.e., valid for arbitrary $\lambda$) result for the three--point function of the ${\mathcal R}$--current operator\footnote{In fact there are also terms proportional to $f^{abc}$. We simply ignore them since they do not contain the epsilon tensor $\epsilon^{\mu\nu\alpha\beta}$. Besides, they vanish if one sets $a=b$.} \cite{Freedman:1998tz,Chalmers:1998xr}
\beq \lim_{y \to 0} \langle J_a^\mu(y) J_b^\nu(0) J_c^\rho(z)\rangle =\frac{N^2}{8\pi^6}d^{abc}\epsilon^{\mu\nu}\,_{\alpha\beta}\frac{y^\alpha }{y^4 (y-z)^6}\left(\eta^{\beta\rho}-2\frac{(y-z)^\beta (y-z)^\rho}{(y-z)^2}\right)\,.  \label{anomaly}
\eeq
This is related to the axial anomaly in the triangle diagram.
In the dual string theory description, it arises from the Chern--Simons term in the effective supergravity action.
Also the two--point function is exactly known.
\beq
\langle J_{c'}^\beta(y) J_c^{\rho}(z) \rangle = \delta_{cc'} \frac{3N^2}{8\pi^4}\frac{1}{(y-z)^6}\left(\eta^{\beta\rho}-2\frac{(y-z)^\beta (y-z)^\rho}{(y-z)^2}\right)\,. \label{jj}
\eeq
One can deduce from (\ref{anomaly}) and (\ref{jj})  an OPE
\beq
J_a^\mu(y) J_b^\nu(0) = d^{abc}\epsilon^{\mu\nu}\,_{\alpha\beta}\frac{y^{\alpha}}{3\pi^2 y^4}J_c^{\beta}(0) + \cdots\,, \label{ope}
\eeq
which  should be valid for all values of $\lambda$. In Appendix A we shall explicitly verify it in the free theory $\lambda=0$. Using (\ref{ope}), one finds
 \beq \int d^4 y \, e^{iqy} \langle PS| T\{J_3^\mu(y) J_3^\nu(0)\}|PS\rangle\Big\arrowvert_{asym} &=& d^{33c}\epsilon^{\mu\nu}\,_{\alpha\beta}\frac{2q^\alpha}{3Q^2}\langle PS| J_c^{\beta}(0)|PS \rangle
 \nonumber \\ &=& d^{33c}  \epsilon^{\mu\nu}\,_{\alpha\beta}\frac{q^{\alpha}}{3P\cdot q}\frac{1}{x}\langle PS| J_c^{\beta}(0)|PS \rangle\,.  \label{33}
 \eeq
 The matrix element has the form
\beq
\langle PS| J_c^{\beta}(0)|PS \rangle
= {\mathcal Q}_c(A S^\beta + BP^\beta)\,, \label{coe}
\eeq
where ${\mathcal Q}_c$ is the ${\mathcal R}$--charge. The first and second terms come from the fermionic and bosonic parts of the ${\mathcal R}$--current, respectively. The coefficients $A$ and $B$ are identified with the elastic form factors at zero momentum transfer. In principle, they can be computed once one specifies an AdS/QCD model. However, an important point is that in order to obtain a nonzero $A$ ($B$ is in general nonzero), it is necessary to employ  models which are dual to the ${\mathcal N}=4$ theory in the UV, and at the same time contain the massless Nambu--Goldstone modes of spontaneously broken ${\mathcal R}$--symmetry. Otherwise, the current conservation law immediately implies $A=0$, as was observed  in a `hardwall' model calculation in \cite{Gao:2009ze}.  An exemplary model which meets this requirement is the one constructed in  \cite{Babington:2003vm}. Keeping such a model in mind, in the following we simply assume that $A\neq 0$. Though this extra assumption is not essential to our paper,  it is in any case useful in order to make some contact with QCD.

(\ref{33}) is just the first term in the OPE, and as such, it  does not contribute to the structure function defined as the imaginary part of the correlator. Rather,
it is related to the first moment of $g_1$ and $F_3$ as in (\ref{g1})
\beq
\int_0^1 dx \,g_1(x,Q^2) = \frac{d^{33c}{\mathcal Q}_c}{12}A\,, \qquad \int_0^1 dx F_3(x,Q^2)=\frac{d^{33c}{\mathcal Q}_c}{6}B\,. \label{sumrule}
\eeq
In QCD, apart from the global constraint (\ref{sumrule}), the first term in the OPE does not tell anything about the actual $x$--dependence of $g_1$. Its small--$x$ behavior has then to be studied quite independently, using the perturbative QCD evolution  \cite{Ball:1995ye,Bartels:1995iu,Bartels:1996wc} or perhaps in the framework of Regge theory  \cite{Heimann:1973hq}. However, in strongly coupled ${\mathcal N}=4$ SYM it turns out that they are closely related. This can be readily understood by looking at the unpolarized case where one takes the energy momentum tensor in the $JJ$ OPE
\beq
JJ \sim \frac{1}{x^2}T\,.
\eeq
The $1/x^2$ dependence can be interpreted as arising from the exchange of a graviton which has spin $j=2$ in the $t$--channel. Once the graviton is Reggeized, a nonzero imaginary part arises, hence the structure function
\beq
F_1(x,Q^2)\sim \left(\frac{1}{x}\right)^{2-{\mathcal O}(1/\sqrt{\lambda})}\,. \label{f11}
\eeq
 In the polarized case, the $1/x$ dependence in (\ref{33}) is due to the $t$--channel exchange of a  Kaluza--Klein photon which is dual to the ${\mathcal R}$--current operator. In the next section, we shall demonstrate that the Reggeization of the photon generates the polarized structure functions which indeed behave as
\beq g_1(x,Q^2), \ F_3(x,Q^2) \sim \left(\frac{1}{x}\right)^{1-{\mathcal O}(1/\sqrt{\lambda})}\,, \label{regge}
\eeq
  in the regime $x\sim e^{-\sqrt{\lambda}}$.

 As for the $g_2(x,Q^2)$ structure function, we  argue that it is much smaller than $g_1$. To see this, note that the derivation of the  relation (\ref{ww}) hinges only on the existence of the OPE and hence it is valid even at strong coupling. Moreover,  the twist--three contribution $\bar{g}_2$ is expected to be negligible because of a large anomalous dimension of order $\sim \lambda^{1/4}$.
 Now suppose that $g_1$ behaves as
\beq
g_1(x,Q^2) \sim \frac{c}{x^{1-\epsilon}}\,,
\eeq
where $\epsilon \sim {\mathcal O}(1/\sqrt{\lambda})$ is a small positive number. Plugging this into (\ref{ww})
 one finds
 \beq
g_2(x)\sim -\frac{c}{1-\epsilon}+\frac{\epsilon c}{(1-\epsilon) x^{1-\epsilon}}\,,
\eeq
 and \beq
  \int_0^1 dx\, g_2(x,Q^2)=0\,.
\eeq
Thus $g_2$ has the same $x$ dependence as $g_1$, but the coefficient is suppressed by a factor of $\epsilon\sim 1/\sqrt{\lambda}$.

We have so far only considered the SU(4) ${\mathcal R}$--currents $J^\mu_a$ that appear in the $JJ$  OPE. On the other hand, from the relation among the SU(4) generators
\beq t^at^b=\frac{\delta^{ab}}{2N_f} +\frac{d^{abc}}{2}t^c + i\frac{f^{abc}}{2}t^c\,, \label{ma}
   \eeq
  with $N_f=4$, it is clear that  the OPE also involves the singlet axial current
\beq J^\mu_{singlet} = \sum_{i=1}^4 \bar{\psi}_i\bar{\sigma}^\mu \psi_i\,, \label{singlet}
\eeq
associated with the U$_{A}$(1) part of the putative U(4)=U$_A$(1)$\times$SU(4) ${\mathcal R}$--symmetry.
 As we mentioned already, the singlet axial current measures the fermions' helicity contribution to the total spin (cf., the $a^{(0)}$ term in (\ref{term})).  The problem with this current in ${\mathcal N}=4$ SYM is that  the U$_A$(1) (or U(4))  is not a symmetry group of the theory, even classically (except in the free theory), due to the Yukawa coupling $\sim f_{abc}\phi_a^{ij}\psi_b^i\psi^j_c$ in the Lagrangian.
 [The scalars cannot have charges under U$_A$(1).] Moreover,
 quantum mechanically, there is an axial anomaly in the divergence of the current
 \beq \partial_\mu J^\mu_{singlet} = -\frac{\lambda}{4\pi^2} F_{\mu\nu}^a\tilde{F}^{\mu\nu}_a +\cdots\,,
 \eeq
 where $\cdots$ denotes the `classical anomaly' from the Yukawa coupling. Due to the strong violation of the `conservation law', the singlet current is not protected under renormalization. The operator (\ref{singlet}) and its superdescendants (the so--called Konishi multiplet) are dual to the string states in the first  excited level. Consequently, the singlet current acquires a large anomalous dimension of order $\lambda^{1/4}$ at strong coupling,\footnote{For a detailed study of the Konishi multiplet and the anomaly in ${\mathcal N}=4$ SYM, see, e.g., \cite{Bianchi:2001cm}. We thank M. Bianchi for answering our questions.}  which makes it completely irrelevant to the present discussion. Thus, as announced in Introduction, we are lead to a drastic conclusion  that the helicity of Weyl fermions does not contribute at all to the total spin for all values of $Q^2>\Lambda^2$.

 What about the gluon helicity contribution $\Delta G$? In the light--cone OPE on which our analysis is based, the singlet current (\ref{singlet}) that appears in the $g_1$ sum rule does not mix with  gluonic operators since there is no local, gauge--invariant spin--1 gluonic operator \cite{Jaffe:1989jz}.  Still, one can define $\Delta G$ as the first moment of a gauge--invariant non--local operator and consider the evolution
 \beq
 \frac{\partial }{\partial \ln Q^2}\begin{pmatrix} \Delta \Sigma \\ \Delta G \end{pmatrix} = \frac{\lambda}{16\pi^2}\begin{pmatrix} -6 & 0 \\ 3 & 0 \end{pmatrix}
 \begin{pmatrix} \Delta \Sigma \\ \Delta G \end{pmatrix}\,,
 \eeq
 where, as an illustration, we show the one--loop result from \cite{Kotikov:2002ab}.
  Unlike in QCD, the (1,1) entry is nonzero because the singlet current is not conserved even classically.
   The anomalous dimension matrix has a left eigenvector
  \beq
  \Delta\Sigma +2\Delta G\,, \label{lin}
  \eeq
  with zero eigenvalue.
Due to superconformal invariance, the linear combination (\ref{lin}) has vanishing anomalous dimension to all orders \cite{Kotikov:2002ab}.
Combining this fact with the anomalous dimension  of  $\Delta \Sigma$ at strong coupling, $\gamma\approx -\lambda^{1/4}$, one finds
\beq
\Delta \Sigma(Q^2) =\tilde{C}\left(\frac{\Lambda^2}{Q^2}\right)^{\lambda^{1/4}}\,, \qquad \Delta G(Q^2)=-\frac{\tilde{C}}{2}\left(\frac{\Lambda^2}{Q^2}\right)^{\lambda^{1/4}}+C\,, \label{helic}
\eeq
where $C$ and $\tilde{C}$ are integration constants (independent of $Q^2$), and therefore,
\beq
\Delta \Sigma(Q^2) + \Delta G(Q^2)=\frac{\tilde{C}}{2}\left(\frac{\Lambda^2}{Q^2}\right)^{\lambda^{1/4}}+C \approx C\,.
\eeq
We see that there is a very rapid transfer of angular momentum from the helicity degrees of freedom to the orbital motion. In practice, these two contributions have constant values $C$ and $\frac{1}{2}-C$, respectively, for all $Q^2>\Lambda^2$.

 Finally in this section, we comment on the physical meaning of the $g_1$ sum rule in the ${\mathcal N}=4$ theory.
  Setting  $a=b=3$ in (\ref{ma}),
 we get
 \beq
 \frac{d^{33c}}{2}t^c=\frac{1}{8}\begin{pmatrix} 1 & 0 & 0 & 0 \\ 0 & 1 & 0 & 0 \\
 0& 0& -1 & 0 \\
 0 & 0& 0& -1 \end{pmatrix}\,.
\label{blo} \eeq
Therefore,\footnote{Note that Wely fermions contain only two degrees of freedom. This is why there are only two terms (as opposed to four, cf., (\ref{heli})) in (\ref{two}) for each flavor. Also, (\ref{two}) is valid if the helicity of the hadron is $h=+\frac{1}{2}$. If $h=-\frac{1}{2}$, then one should rather write
\beq g_1(x)
 \sim \sum_{i=1,2} (\bar{\psi}_i^+(x)-\psi_i^-(x)) -\sum_{i=3,4} (\bar{\psi}_i^+(x)-\psi_i^-(x))\,.
 \eeq
}
 \beq g_1(x)
 \sim \sum_{i=1,2} (\psi_i^+(x)-\bar{\psi}_i^-(x)) -\sum_{i=3,4} (\psi_i^+(x)-\bar{\psi}_i^-(x))+ ({\rm bosonic \ contributions})\,.
 \label{two}
 \eeq
Thus the $g_1$ sum rule (\ref{sumrule}) is a direct analog of the Bjorken sum rule (\ref{bjo}) (no need to subtract the singlet part since there is none) which simply represents the conservation of the axial, or rather the ${\mathcal R}$--charge corresponding to the generator (\ref{blo}).



\section{The $g_1$ structure function at strong coupling}
The goal of this section is to derive
 (\ref{regge}) in its precise form. We first consider an approach based on the string S--matrix in AdS$_5$ following \cite{Brower:2006ea} (see also, \cite{Hatta:2007he,Brower:2007xg,Brower:2008cy}). Based on this, we then develop a pragmatic approach suited for our problem and construct the $g_1$ structure function which exactly satisfies the sum rule (\ref{sumrule}).

\subsection{Reggeized photon exchange in AdS}
 Scattering amplitudes in type IIB string theory in the background of AdS$_5\times$S$^5$ show the usual Regge behavior at small momentum transfer $\tilde{s}\gg |\tilde{t}|$
\beq {\mathcal A} \sim \tilde{s}^{2+\frac{\alpha'\tilde{t}}{2}}\,, \label{grav}
\eeq
 where we put a tilde on the Mandelstam variables in order to distinguish them from the corresponding quantities in the four--dimensional boundary gauge theory.
The $\tilde{s}^2$--dependence is due to the exchange of the (Reggeized) graviton $G_{++}$ which has spin $j=2$ in the light--cone direction. On the other hand, the photon  associated with  the
 ${\mathcal R}$--current operator is, from the ten--dimensional point of view, a component of the graviton with one of its indices polarized in the S$^5$ direction
\beq
 G_{+a}=A_+K_a\,,
 \eeq
 where $K_a$ ($a=1,2,3,4,5$) is a Killing vector. Since the momenta in the S$^5$ direction carried by the external states are at most of order $1/R$ ($R=1$ is the common radius of AdS$_5$ and S$^5$), the photon exchange gives an amplitude smaller than (\ref{grav}) by a factor of $\tilde{s}$
\beq   {\mathcal A} \sim \tilde{s}^{1+\frac{\alpha'\tilde{t}}{2}}\,. \label{photo}
 \eeq
 As suggested in \cite{Polchinski:2002jw,Brower:2006ea}, the momentum transfer $\tilde{t}$ in (\ref{grav}) and (\ref{photo}) can be interpreted as the Laplacian operator acting on the eight--dimensional transverse space. The form of the Laplacian is determined by looking at the equation of motion obeyed by the field exchanged in the $t$--channel. Generically, it takes the form
\beq
(\Delta_j+f(j))\Phi_{j+} =0\,. \label{ki}
\eeq
for a spin--$j$ field
\beq
\Phi_{j+}\equiv \Phi_{\scriptsize{\underbrace{++\cdots +}_{j\  {\rm indices}}}}\,.
\eeq
 In (\ref{ki})   $f(j)$ is a c--number function  and
the differential operator $\Delta_j$ is defined as
\beq \Delta_j \equiv \frac{1}{z^{j}}\Delta_0 z^j\,, \eeq
 where $\Delta_0$ is the scalar Laplacian in AdS$_5$
 \beq
\Delta_0=-z^2(\partial_z^2+\partial_\perp^2)+3z\partial_z\,.
\eeq
Note that we keep only the derivative $\partial_\perp^2 \equiv \partial_{y_1}^2+\partial_{y_2}^2$ in the transverse direction of the Minkowski space. At high energy this factor is essentially identified with the momentum transfer in the boundary gauge theory $\partial_\perp^2 \to t$.

 For example, the graviton  $G_{++}$  obeys the linearized Einstein equation in AdS$_5$ which reduces to
 \beq
\Delta_2 G_{++} =(-z^2\partial_z^2 - z^2 \partial_\perp^2 - z\partial_z + 4)G_{++}=0\,.
\eeq
   Thus one sees that $f(2)=0$. The amplitude (\ref{grav}) then becomes an operator
   \beq {\mathcal A} \sim \tilde{s}^{2-\alpha'\Delta_2/2}\,,
\eeq
 acting on the hyperbolic space H$_3$ parameterized by $(z,\vec{y}_\perp$).
   Going to the basis which diagonalizes  the operator $\Delta_2$, one finds that
   \beq
   {\mathcal A} \sim \tilde{s}^{2-2/\sqrt{\lambda}}\,. \label{bev}
   \eeq
   Thus the graviton intercept is slightly shifted from 2 due to the curvature of AdS$_5$, and this can be identified with the Pomeron intercept \cite{Kotikov:2004er,Brower:2006ea}.
  In unpolarized DIS, (\ref{bev}) is essentially the origin the behavior (\ref{f11}) due to the correspondence $\tilde{s} \leftrightarrow 1/x$.

   Similarly, the photon field obeys the five--dimensional Maxwell equation
 \beq D_M F^{M -} &=&z^2(-z^2\partial_z^2
  - z^2 \partial_\perp^2+z\partial_z ) A_+  \nonumber \\
  &=&z^2 (\Delta_1 - 3)A_+=0\,. \eeq
 Thus  $f(1)=-3$ in this case, meaning that the field  $zA_+$ behaves as a scalar with negative mass squared $m^2=-3$.
The amplitude (\ref{photo}) becomes
 \beq {\mathcal A} \sim \tilde{s}^{1-\alpha'(\Delta_1-3)/2}\,. \label{amp}
\eeq
Though it is straightforward to diagonalize the operator $\Delta_1$, in order to make contact with the OPE (\ref{inf}) it is convenient at this point to express (\ref{amp}) as a contour integral in the complex $j$--plane
 \beq
 \langle u| \tilde{s}^{1-\alpha'(\Delta_1-3)/2}|u'\rangle &=&\int \frac{dj}{2\pi i} \langle u | \frac{\tilde{s}^j}{j-1+\alpha'(\Delta_j-3)/2}|u'\rangle \nonumber \\
 &=& 2\sqrt{\lambda} \int \frac{dj}{2\pi i}\tilde{s}^j \, \sqrt{G'}(G'^{+-})^jD_{j+j-}(u,u')\,, \label{un}
\eeq
where we collectively denoted $u\equiv (z,y^\mu)$ and used the relation $\alpha'=1/\sqrt{\lambda}$ which holds in the present units. In the second equality, we have
   defined the $t$--channel propagator $D_{j+j-}$ of the photon whose spin is analytically continued away from $j=1$. It satisfies the following relation
\beq
 \left(\Delta_j-3+2\sqrt{\lambda}(j-1)\right)D_{j+j-}(u,u')=
 \frac{\delta^{(5)}(u-u')}{(G'^{+-})^j\sqrt{G'}}\,. \label{prop}  \eeq
    The solution is, for $t=\partial^2_\perp=0$,
 \beq
 D_{j+j-}(u,u')=(zz')^{2-j}\int \frac{d\nu}{\pi} \frac{e^{-i\nu(\rho-\rho')}}{4\nu^2+1+2\sqrt{\lambda}(j-1)}\delta^{(4)}(y-y')\,,
 \eeq
  where we have defined $z^2=e^{-\rho}$. Then (\ref{un}) can be evaluated as
 \beq
 \langle u| \tilde{s}^{1+\alpha'(\Delta_1+3)/2}|u'\rangle &\approx & \tilde{s}^{j_A}(zz')^{2-j_A}\int \frac{d\nu}{\pi} \tilde{s}^{-D\nu^2}
  e^{-i\nu(\rho-\rho')}\delta^{(4)}(y-y')\nonumber \\  &= & \tilde{s}^{j_A}(zz')^{2-j_A}
 \frac{e^{-\frac{(\rho-\rho')^2}{4D\tau}}}{\sqrt{\pi D\tau}} \delta^{(4)}(y-y')\,, \label{fin}
 \eeq
 where $D\equiv 2/\sqrt{\lambda}$ is the `diffusion' parameter, and $\tau\equiv \ln \tilde{s}$ is the rapidity. The Reggeized photon intercept
 \beq
 j_A \equiv 1-\frac{1}{2\sqrt{\lambda}}\,, \label{reg}
 \eeq
  is slightly shifted from 1 due to the nonzero curvature of AdS$_5$.
(\ref{fin})  implies that the structure functions indeed exhibit the behavior anticipated in (\ref{regge}). It also shows that the interaction is nonlocal  in the fifth dimension due to diffusion: In a complete calculation  (\ref{fin}) should be convoluted with the target wavefunctions localized around $z\sim 1/\Lambda$ and the ($s$--channel) photon wavefunctions localized around $z'\sim 1/Q$.

\subsection{A pragmatic approach}
One can incorporate the procedures in the previous subsection into the worldsheet OPE approach \cite{Brower:2006ea,Brower:2008cy} by introducing vertex operators for the external and internal states. However, applied to our problem of polarized DIS,  it appears difficult in this approach to obtain the expected tensor structure (namely, the epsilon tensor $\epsilon^{\mu\nu\alpha\beta}$ in (\ref{tens})) and determine the normalization factor essential to discuss the sum rule. We therefore take a different and more pragmatic tack. The idea is to  try to reconstruct the imaginary part from the first term in the OPE (\ref{33}). Let us focus on the factor
\beq
\frac{1}{x}\langle PS| J_c^{+}(0)|PS \rangle\,. \label{mat}
\eeq
 The insertion of the operator $J_c^+(0)$ excites a non--normalizable mode of the bulk gauge field
 \beq A_+(z,y)=\frac{2i}{\pi^2}\int d^4y'\frac{z^2}{(z^2-(y-y')^2+i\epsilon)^3}A_+(y')\,, \label{kern} \eeq
  where $A_+(x')$ is the boundary source which, for a local insertion $J^+_c(y'=0)$, is given by a delta function   $A_\mu(y')=\delta_\mu^+\delta^{(4)}(y')$.
  The integral kernel in (\ref{kern}) is the bulk--to--boundary propagator.  The matrix element (\ref{mat}) may be evaluated as
  \beq
  \frac{1}{x}\langle PS| J_c^{+}(0)|PS \rangle=\frac{{\mathcal Q_c}}{x} \int d^4y \int dz \sqrt{G} A_+(z,y)\bar{\psi}\gamma^+\psi(z,y)\,. \label{dis}
  \eeq
  The $y$--dependence of the target wavefunctions is simply the plane wave $e^{iP\cdot y}$ which cancels between $\psi$ and $\bar{\psi}$ (for the forward scattering). The $d^4y$ integral can then be done
  \beq \int d^4y A_+(z,y)=\frac{2i}{\pi^2}\int d^4y d^4y'\frac{z^2}{(z^2-(y-y')^2+i\epsilon)^3}A_+(y')
  =\int d^4y' A_+(y')=1\,, \eeq
  so that (\ref{dis}) reduces to
   \beq
  \frac{1}{x}\langle PS| J_c^{+}(0)|PS \rangle=\frac{{\mathcal Q_c}}{x}  \int dz \sqrt{G} \bar{\psi}\gamma^+\psi(z) =\frac{{\mathcal Q}_c}{x}(AS^+ + BP^+)\,.
  \eeq

  Alternatively, one can introduce an effective bulk current $J_{bulk}$ which sources (\ref{kern}) via the bulk--to--bulk propagator.
  \beq A_+(z,y)=\int d^4y'dz' \sqrt{G'}D_{+-}(u,u')J^-_{bulk}(z',y')\,. \label{one} \eeq
   Naturally, $J_{bulk}$ has a support at small $z<1/Q$ reflecting the uncertainty $|\Delta y^2|\sim 1/Q^2$ in the location of the operator insertion (\ref{mat}).
 Using the asymptotic form of the propagator \cite{D'Hoker:1998gd} (adapted to the Minkowski metric)
\beq D_{+-}(u,u')\approx \frac{i}{\pi^2 zz'}\left(\frac{zz'}{z^2-(y-y')^2+i\epsilon}\right)^3\,,  \qquad (z'\to 0) \eeq
one finds the following relation between $J^{bulk}_+(z,y)$ and $A_+(y)$
 \beq
 A_+(y')=\frac{1}{2}\int \frac{dz'}{z'}J_+^{bulk}(z',y')\,. \label{ide}
 \eeq
 [Note that $J^-=z'^2J_+$.]
The propagator $D_{+-}$ is identical to the one defined in (\ref{prop}) with $j=1$.
 (\ref{one}) can be rewritten as
\beq
A_+(z,y)=\int d^5u' \frac{1}{\Delta_1-3}\delta^{(5)}(u-u')J_+^{bulk}(u')\,. \label{become}
\eeq

\begin{figure}
\begin{center}
\centerline{\epsfig{file=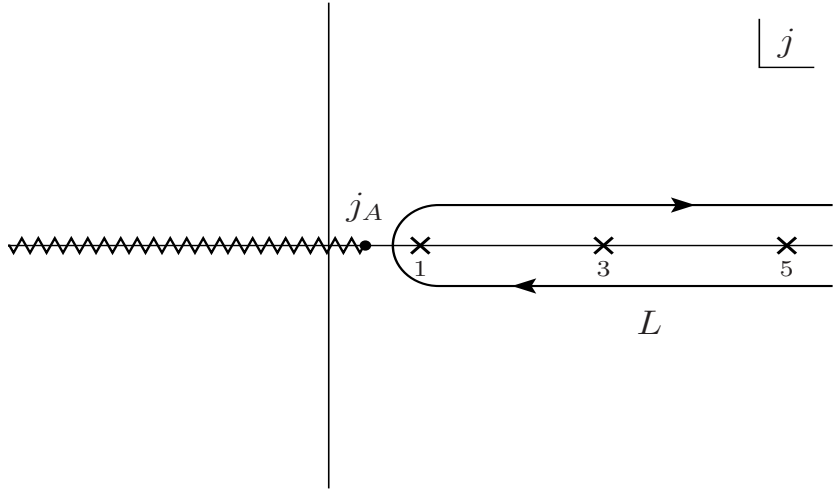,height=6.5cm,width=11.cm}}
\caption{\sl
The integration contour in the complex $j$--plane. A branch cut appears after integrating over $\nu$.
\label{contour} }
\end{center}
\end{figure}

So far we have dealt with only a single term in the $JJ$ OPE. However, our experience with QCD tells us that one should sum over the twist--two operators with odd $j$ values as in (\ref{inf}). We do this in the form of a contour integral in the complex $j$--plane
\beq
\sum_j^{odd} = \int_L \frac{dj}{4i}\frac{1-e^{-i\pi j}}{\sin \pi j}\,, \label{cont}
\eeq
 where the contour $L$ is shown in fig.~\ref{contour}.
 Accordingly, all the factors in (\ref{dis}) should be analytically continued  away from $j = 1$. As for the propagator, it suffices to use (\ref{prop}) in anticipation  that in the end the important values of $j$ will be close to $j=1$.
 (\ref{become}) is then replaced by
 \beq
 A_{j+}(z,y)=\int d^5u'\frac{1}{\Delta_j-3+2(j-1)/\alpha'}\delta^{(5)}(u-u')J_{j+}^{bulk}(u')\,.
 \label{a+} \eeq
On the other hand, the factor $1/x$ obviously generalizes to
\beq \frac{1}{x} \to \left(\frac{1}{x}\right)^j\,. \label{x+} \eeq
All in all, (\ref{dis}) becomes\footnote{There may be an unknown, multiplicative factor $c(j)$ in the integrand of (\ref{all}) such that $c(1)=1$. However, the $j$--integral will be dominated by the region $j\approx 1$ where the function $c(j)$ must be slowly varying.  [Physically, we do not see any source of rapid variation in $j$ near $j=1$ other than (\ref{a+}) and (\ref{x+}).] Therefore we may set $c(j)=1$ from the beginning.}
\beq
I&\equiv &{\mathcal Q_c}\int \frac{dj}{4i}\frac{1-e^{-i\pi j}}{\sin \pi j}\left(\frac{1}{x}\right)^j\int d^4y dz \sqrt{G}\int d^4y' dz' \nonumber \\ && \times \frac{1}{\Delta_j-3+2(j-1)/\alpha'}\delta^{(5)}(u-u')J_{j+}^{bulk}(u') \bar{\psi}\gamma^+(\partial^+)^{j-1}\psi(z)\,. \label{all}
\eeq
Let us do the $j$-integral first. Deforming the contour to pick up the pole from the propagator (see, fig.~\ref{contour}), we get \footnote{To avoid cumbersome expressions, we retain the letter $j$ in (\ref{j}) as representing the pole of the propagator: $j=1-\alpha'(\Delta_j-3)/2$.}
\beq
I&=&{\mathcal Q_c}\frac{\pi}{4\sqrt{\lambda}}\int d^4y dz \sqrt{G}\int d^4y'dz'\frac{1-e^{-i\pi j}}{\sin \pi j}\left(\frac{1}{x}\right)^{1-\alpha'\frac{\Delta_j-3}{2}}  \nonumber \\
&& \times \delta^{(5)}(u-u')J_{+}^{bulk}(u') \bar{\psi}\gamma^+\psi(z)\,. \label{j}
\eeq
 In order to diagonalize the Laplacian
 operator, we  notice that
 \beq
 \left(\frac{1}{x}\right)^{1-\alpha'\frac{\Delta_j-3}{2}}=z^{2-j}\left(\frac{1}{x}\right)^{1-
 \alpha'\frac{\Delta_2-3}{2}}
 z^{j-2}\,,
 \eeq
  where
  \beq -\Delta_2+3=z^2 \partial_z^2+ z\partial_z + z^2\partial_\perp^2-1=4\partial_\rho^2+z^2t-1\,.
  \eeq
In the case of forward scattering $t=0$, the eigenfunction of the operator $\Delta_2$ is simply a plane wave in $\rho$, so the following representation of the delta function is useful
\beq \delta(z-z')=\int \frac{d\nu}{\pi}\left(\frac{1}{z}\right)^{j-2+2i\nu}z'^{j-3+2i\nu}
=\int \frac{d\nu}{\pi}\left(\frac{1}{z}\right)^{j-2}z'^{j-3}e^{i\nu(\rho-\rho')}\,.
\eeq
 (\ref{j}) can then be evaluated as
 \beq
 I&\approx &{\mathcal Q_c}\frac{\pi}{4\sqrt{\lambda}}\int d^4y dz \sqrt{G} \int dz' \sqrt{G'}(G'^{+-})^{j_A}(zz')^{2-j_A} \nonumber \\
 &&  \qquad \quad \times \frac{1-e^{-i\pi j_A}}{\sin \pi j_A}\int \frac{d\nu}{\pi} \left(\frac{1}{x}\right)^{j_A-\frac{2\nu^2}{\sqrt{\lambda}}}e^{i\nu (\rho-\rho')} J_{+}^{bulk}(z',y) \bar{\psi}\gamma^+\psi(z) \label{imag}\\
 &\approx & {\mathcal Q_c}\left(1+i\frac{\pi}{4\sqrt{\lambda}} \right)\int d^4y dz \sqrt{G} \int dz' \sqrt{G'} (G'^{+-})^{j_A}  (zz')^{2-j_A} \nonumber \\
 &&  \qquad \quad \times  \left(\frac{1}{x}\right)^{1-\frac{1}{2\sqrt{\lambda}}} \frac{e^{-(\rho-\rho')^2/4D\tau}}{\sqrt{\pi D \tau}}J_{+}^{bulk}(z',y) \bar{\psi}\gamma^+\psi(z)\,,
 \nonumber
 \eeq
 where $j_A=1-1/2\sqrt{\lambda}$ is as in (\ref{reg}) and $\tau=\ln 1/x$. We remind the reader that the calculation is valid in the regime $x\sim e^{-\sqrt{\lambda}}$, or $(1/x)^{1/\sqrt{\lambda}} \sim {\mathcal O}(1)$.
 In (\ref{imag}), one recognizes an imaginary part which can be identified with the $g_1$ structure function. Working out the prefactor, we arrive at\footnote{Since we keep a small imaginary part of order $1/\sqrt{\lambda}$, consistency requires that one should neglect the ${\mathcal O}(1/\sqrt{\lambda})$ terms in $(G'^{+-})^{j_A}(zz')^{2-j_A}$.}
\beq
g_1(x,Q^2)S^+ + \frac{F_3(x,Q^2)}{2}P^+&=&\frac{d^{33c}{\mathcal Q}_c}{6\pi } \frac{\pi}{4\sqrt{\lambda}}\int d^4y dz \sqrt{G} \int dz' \sqrt{G'} G'^{+-}  zz' \nonumber \\
 && \times  \left(\frac{1}{x}\right)^{1-\frac{1}{2\sqrt{\lambda}}} \frac{e^{-(\rho-\rho')^2/4D\tau}}{\sqrt{\pi D \tau}}J_{+}^{bulk}(z',y) \bar{\psi}\gamma^+\psi(z)\,. \label{find}
\eeq
 This is the main result of this paper. The $Q^2$--dependence is implicit in $J^{bulk}_+$. It is important to notice that the Regge intercept $j_A$ is independent of $Q^2$.

Let us check that (\ref{find}) satisfies  the sum rule.
For this purpose it is convenient to undo the $\nu$--integral and integrate over  $x$ (or equivalently, $\tau$) first. Using
 \beq &&\int \frac{d\nu}{\pi} e^{i\nu (\rho-\rho')}\int_0^\infty d\tau e^{-\tau + (1-1/2\sqrt{\lambda})\tau-2\nu^2\tau/\sqrt{\lambda}} \nonumber \\
 &&= \frac{\sqrt{\lambda}}{2}\int_{-\infty}^\infty \frac{d\nu}{\pi}e^{i\nu(\rho-\rho')}
 \frac{1}{\nu^2+\frac{1}{4}}\nonumber \\
 &&=\sqrt{\lambda}\frac{z'}{z}\,, \eeq
 where $z'\sim 1/Q <z \sim 1/\Lambda$,
 we get
\beq \int_0^1 dx g_1(x,Q^2)&=&\frac{d^{33c}{\mathcal Q}_c}{6\pi S^+} \frac{\pi}{4\sqrt{\lambda}}\sqrt{\lambda}\int d^4ydz \sqrt{G} \int \frac{dz'}{z'}J_{+}^{bulk}(z',y) \bar{\psi}\gamma^+\psi(z)
\nonumber \\ &=&\frac{d^{33c}{\mathcal Q}_c}{12S^+} \int dz \sqrt{G}\bar{\psi}\gamma^+\psi(z)
\nonumber \\ &=&\frac{d^{33c}{\mathcal Q}_c}{12}A\,,
 \eeq where we take only the term proportional to $S^+$ in the integral. In the second equality, we used (\ref{ide}). This is exactly (\ref{sumrule}).

It is also instructive  to identify the region of $x$ which dominantly contributes to the sum rule.
Doing the $\nu$ integral first, we encounter the following integral
\beq
\int_0^1 dx \,g_1(x,Q^2)\sim
 \int_0^\infty d\tau e^{-\frac{\tau}{2\sqrt{\lambda}} -\frac{\ln^2 z^2/z'^2}{4D\tau}}\,. \eeq
 The integrand is strongly peaked at
 \beq
 \tau= \frac{\sqrt{\lambda}}{2}\ln \frac{z^2}{z'^2} \sim \frac{\sqrt{\lambda}}{2}\ln \frac{Q^2}{\Lambda^2}\,. \label{val}
 \eeq
According to our interpretation of the $g_1$ structure function, (\ref{val}) is the value  around which the ${\mathcal R}$--charge of the target is concentrated. Incidentally, we note that
 the energy--momentum sum rule is saturated around the value
\cite{Hatta:2007he}\beq
 \tau=  \frac{\sqrt{\lambda}}{4}\ln \frac{Q^2}{\Lambda^2}\,.
 \eeq

Finally, consider the anomalous dimension of twist--two operators.
This can be read off from the $z'$ dependence of the integrand.
In (\ref{imag}) one recognizes the factor
\beq
\left(\frac{1}{z'}\right)^{j-2}e^{-i\nu \rho'}=\left(\frac{1}{z'}\right)^{j-2-2i\nu}=\left(\frac{1}{z'^2}\right)^{\gamma(j)-1}
\sim (Q^2)^{\gamma(j)-1}\,,
\eeq
where \cite{Brower:2006ea,Hatta:2007he}
\beq \gamma(j)=\frac{j}{2}-i\nu\,. \eeq
This means that
\beq \left(\frac{j}{2}-\gamma\right)^2=-\nu^2=\frac{1}{D}(j-j_A)\,, \label{means}
\eeq
or equivalently,
\beq
\gamma(j)=\frac{j}{2}-\sqrt{\frac{\sqrt{\lambda}}{2}\left(j-1+\frac{1}{2\sqrt{\lambda}}\right)}\,.
\eeq
 The total dimension is then
 \beq \Delta(j)=j+2-2\gamma(j)=2+2\sqrt{\frac{\sqrt{\lambda}}{2}\left(j-1+\frac{1}{2\sqrt{\lambda}}\right)}\,. \label{tot}
 \eeq
 When $j=1$, $\Delta(1)=3$ and $\gamma(1)=0$. This is the ${\mathcal R}$--current operator which is protected.
 The above formula gives an analytic continuation to $j>1$. Inverting (\ref{tot}), one gets an expression
 \beq
 j=j_A + \frac{(\Delta-2)^2}{2\sqrt{\lambda}}\,,
 \eeq
 which clearly shows the the known symmetry $\Delta \leftrightarrow 4-\Delta$.
Also, it is interesting to note that one can write
 \beq
 \Delta = 2\pm \sqrt{1+m^2}\,, \label{vector}
 \eeq
 with
 \beq m^2=\frac{2(j-1)}{\alpha'}\,. \label{mass}
 \eeq
 (\ref{vector}) is the formula  which relates the dimension of  field theory operators and the mass of vector fields in AdS$_5$, and (\ref{mass}) is  the mass of the spin $j=2n+1$ ($n=0,1,2,\cdots$) state in the $n$--th excited level $m^2=4n/\alpha'$ of type IIB superstring theory.

\section{Discussions}

 \subsection{`Asymptotic' spin decomposition}
  It was observed by Ji \cite{Ji:1996ek} that in QCD the partition of the proton spin among quarks and gluons approaches the ratio $3N_f:16$ in the asymptotic limit $Q^2 \to \infty$. Here the numbers include both the helicity and orbital contributions. The result has been obtained quite independently of the parton dynamics in DIS, and is solely based on the renormalization of the angular momentum operator which turns out to be identical to that of the energy--momentum tensor.  In ${\mathcal N}=4$ SYM one can repeat the same analysis. The  energy--momentum operator is given as a sum of contributions from gluons, Weyl fermions and scalars
\beq
T_{\mu\nu}=T_{\mu\nu}^g+T_{\mu\nu}^\psi+ T_{\mu\nu}^\phi\,.
\eeq
There are two other operators which are eigenvectors of the twist--two anomalous dimension matrix with spin $j=2$
\beq
T'=-2T^g+T^\psi +2T^\phi\,,
\nonumber \\
T''=-T^g + 4T^\psi -6T^\phi\,.
\eeq
The coefficients are actually exact, i.e., they do not receive radiative corrections \cite{Anselmi:1998ms,Kotikov:2002ab}. [See, also, Appendix C of \cite{Hatta:2007cs}.] Therefore, for all values of the coupling one can decompose as
\beq
T^g=\frac{2}{5}T-\frac{2}{7}T'-\frac{1}{35}T''\,, \nonumber \\
T^\psi=\frac{2}{5}T+\frac{1}{7}T' + \frac{4}{35}T''\,, \nonumber \\
T^\phi=\frac{1}{5}T+\frac{1}{7}T'-\frac{3}{35}T''\,.
\eeq
This implies that both the total energy  and the total spin are partitioned among gluons, fermions and scalars with the ratio $2:2:1$. A novel feature at strong coupling is that $T'$ and $T''$ acquire a large anomalous dimension of order $\lambda^{1/4}$. Therefore, the ratio is realized not only asymptotically, but in practice for all values of $Q^2$.

Let us comment on a further decomposition into the helicity and orbital parts. Of course the scalars $\phi$  have only orbital angular momentum. As we have argued, the contribution of the  Weyl fermions purely consists of orbital angular momentum. Although $\Delta G$  remains undetermined in our approach,\footnote{Note that in Ji's approach the gluon spin cannot be further decomposed into helicity and orbital parts in a gauge invariant way. The gluon helicity contribution $\Delta G$ we have been discussing is the first moment of a certain nonlocal gauge--invariant operator. Then the orbital part is defined as the rest.} we are inclined to think that the gluons  give only orbital angular momentum as well, i.e., $C=0$ in (\ref{helic}), since we see no reason why the gluons are special. Thus we conclude that the strong coupling effects tend to suppress the helicity contribution, and in the limit of large coupling the entire hadron spin is of orbital origin.

  \subsection{ Small--$x$ behavior}
  Let us compare our result in ${\mathcal N}=4$ SYM at small--$x$
  \beq
  g_1^S(x) \approx 0, \qquad g_1^{NS}(x) \sim \left(\frac{1}{x}\right)^{1-1/2\sqrt{\lambda}}\,, \label{our}
  \eeq
  for the singlet and non--singlet parts, respectively, with the corresponding behavior  in QCD. Experimentally, within the currently accessible range of $x$ the flavor non--singlet contribution to $g_1$ rises as $x^{-\alpha}$ with $\alpha\gtrsim 0.5$ \cite{Abe:1997cx,Bass:2006dq} whereas the flavor singlet contribution is almost zero at small--$x$ \cite{Alexakhin:2006vx}. This is qualitatively similar to (\ref{our}), though of course such a comparison should be taken with caution.
   On the other hand, predictions from perturbative QCD vary \cite{Ball:1995ye,Bartels:1995iu,Bartels:1996wc} depending on which terms in higher order graphs are resummed. Including the so--called double logarithmic contributions, Refs.~\cite{Bartels:1995iu,Bartels:1996wc} obtained
   \beq
   g_1^S(x)\sim \left(\frac{1}{x}\right)^{k\frac{\sqrt{\lambda}}{2\pi}}, \qquad g_1^{NS}(x)\sim \left(\frac{1}{x}\right)^{\frac{\sqrt{\lambda}}{2\pi}}\,, \label{hing}
   \eeq
    with $k\approx 2.5$, which clearly means $|g_1^{S}|\gg |g_1^{NS}|$ at low--$x$. Whether this opposite behavior can be observed in a yet unexplored region of $x$ in future experiments  is not clear at the moment. In any case, it is interesting to notice that the exponent of the non--singlet part appears to have a smooth interpolation from 0 to 1 as the coupling $\lambda$ is varied from zero to infinity.
   We finally mention that in the conventional Regge theory, the singlet and non--singlet parts are governed by the $f_1$ and $a_1$ meson Regge trajectories with \emph{negative} intercepts  $-0.18 \gtrsim \alpha_{a_1}, \, \alpha_{f_1}\gtrsim -0.4$. Therefore, $g_1(x)$ goes to zero as $x\to 0$ in this framework.\\

\section*{Acknowledgments}
We thank Shunzo Kumano and Werner Vogelsang for useful correspondence regarding the uncertainty in  $\Delta G$.
The work of Y.~H. and T.~U. is supported, in part, by Special Coordination Funds for Promoting Science and Technology of the Ministry of Education, Culture, Sports, Science and Technology, the Japanese Government. B.~X. is supported  by the Director, Office of Energy
Research, Office of High Energy and Nuclear Physics, Divisions of
Nuclear Physics, of the U.S. Department of Energy under Contract No.
DE-AC02-05CH11231.

\appendix

\section{OPE at weak coupling}
Since (\ref{ope}) is  an exact result for any value of the coupling, one should be able to check it in the free theory. The ${\mathcal R}$--current operator is
\beq J^\mu_a&=& \bar{\psi}_i\bar{\sigma}^\mu t^a_{ij} \psi_j +i\phi_{ij}^\dagger(\partial^\mu -\overleftarrow{\partial}^\mu) T^a_{ij,kl}\phi_{kl} \nonumber \\
&= &\bar{\psi}\gamma^\mu  \frac{1+\gamma_5}{2}t^a\psi +i\phi_{ij}^\dagger(\partial^\mu -\overleftarrow{\partial}^\mu) T^a_{ij,kl}\phi_{kl}\,,
\eeq
 where  $i,j=1,2,3,4$ and in the second line we use the Majorana four--component representation.
For $a=3$,
\beq J^\mu_3= \frac{1}{2}\left(\bar{\psi}^1\bar{\sigma}^\mu \psi^1 - \bar{\psi}^2 \bar{\sigma}^\mu \psi^2\right)+\frac{i}{2}\left(\phi_{13}^\dagger\partial^\mu \phi_{13}-\partial^\mu \phi_{13}^\dagger\phi_{13}\right) +\frac{i}{2}\left(\phi_{14}^\dagger\partial^\mu \phi_{13}-\partial^\mu \phi_{14}^\dagger\phi_{14}\right)\,. \nonumber\\ \label{use}  \eeq
In the $JJ$ OPE, the antisymmetric tensor comes solely from the fermionic term. One finds
\beq J^\mu_a(y)J^\nu_b(0)= \epsilon^{\mu\nu\alpha\beta} d^{abc}\frac{y_\alpha}{2\pi^2y^4}\bar{\psi}\gamma_\beta\frac{1+\gamma_5}{2}t^c \psi + \cdots\,.
\eeq
Naively, the coefficient does not match (\ref{ope}). However one has to take into account the operator mixing. This means that one should decompose in the following way
\beq
\bar{\psi}\gamma_\beta\frac{1+\gamma_5}{2}t^c \psi &=&
\frac{2}{3}\left(\bar{\psi}\gamma_\beta\frac{1+\gamma_5}{2}t^c \psi +
i\phi^\dagger(\partial_\beta -\overleftarrow{\partial}_\beta) T^c\phi \right) \nonumber \\
&&+\frac{1}{3}\left(\bar{\psi}\gamma_\beta\frac{1+\gamma_5}{2}t^c \psi -2
i\phi^\dagger(\partial_\beta - \overleftarrow{\partial}_\beta) T^c\phi \right)
\equiv \frac{2}{3}J_\beta^c + \frac{1}{3}\tilde{J}_\beta^c\,,
\eeq
which gives
\beq J^\mu_a(y)J^\nu_b(0)= \epsilon^{\mu\nu\alpha\beta} d^{abc}\frac{y_\alpha}{3\pi^2y^4}\left(J_\beta^c(0)+\frac{1}{2}\tilde{J}^c_\beta(0)\right)+ \cdots\,.
\eeq
The prefactor in front of the ${\mathcal R}$--current precisely reproduces the coefficient in (\ref{ope}). The second term is the orthogonal operator to the ${\mathcal R}$--current operator in the sense that
\beq
\langle J_a^\alpha(y)\tilde{J}_b^\beta(0)\rangle =0\,. \label{or}
\eeq
 To see this, we only need to note that the coefficient in (\ref{jj}) decomposes as $\frac{3}{8}=\frac{1}{4}+\frac{1}{8}$ where $\frac{1}{4}$ and $\frac{1}{8}$ are the fermionic and  bosonic contributions, respectively.   At strong coupling, $\tilde{J}$ acquires a large anomalous dimension, so it is irrelevant.

\bibliographystyle{utcaps}
\bibliography{polreference}

\providecommand{\href}[2]{#2}\begingroup\raggedright\begin{thebibliography}{10}

\bibitem{Ashman:1987hv}
{\bf European Muon} Collaboration, J.~Ashman {\em et al.}, ``{A measurement of
  the spin asymmetry and determination of the structure function g(1) in deep
  inelastic muon proton scattering},'' {\em Phys. Lett.} {\bf B206} (1988)
364.

\bibitem{Kuhn:2008sy}
S.~E. Kuhn, J.~P. Chen, and E.~Leader, ``{Spin Structure of the Nucleon -
  Status and Recent Results},''
\href{http://arXiv.org/abs/0812.3535}{{\tt 0812.3535}}.

\bibitem{Alexakhin:2006vx}
{\bf COMPASS} Collaboration, V.~Y. Alexakhin {\em et al.}, ``{The Deuteron
  Spin-dependent Structure Function g1d and its First Moment},'' {\em Phys.
  Lett.} {\bf B647} (2007) 8--17,
\href{http://arXiv.org/abs/hep-ex/0609038}{{\tt hep-ex/0609038}}.

\bibitem{Leader:2006xc}
E.~Leader, A.~V. Sidorov, and D.~B. Stamenov, ``{Impact of CLAS and COMPASS
  data on Polarized Parton Densities and Higher Twist},'' {\em Phys. Rev.} {\bf
  D75} (2007) 074027,
\href{http://arXiv.org/abs/hep-ph/0612360}{{\tt hep-ph/0612360}}.

\bibitem{Abelev:2007vt}
{\bf STAR} Collaboration, B.~I. Abelev {\em et al.}, ``{Longitudinal
  double-spin asymmetry for inclusive jet production in p+p collisions at
  sqrt(s)=200 GeV},'' {\em Phys. Rev. Lett.} {\bf 100} (2008) 232003,
\href{http://arXiv.org/abs/0710.2048}{{\tt 0710.2048}}.

\bibitem{Adare:2008px}
{\bf PHENIX} Collaboration, A.~Adare {\em et al.}, ``{The polarized gluon
  contribution to the proton spin from the double helicity asymmetry in
  inclusive $\pi^0$ production in polarized p+p collisions at sqrt(s)=200
  GeV},''
\href{http://arXiv.org/abs/0810.0694}{{\tt 0810.0694}}.

\bibitem{deFlorian:2008mr}
D.~de~Florian, R.~Sassot, M.~Stratmann, and W.~Vogelsang, ``{Global Analysis of
  Helicity Parton Densities and Their Uncertainties},'' {\em Phys. Rev. Lett.}
  {\bf 101} (2008) 072001,
\href{http://arXiv.org/abs/0804.0422}{{\tt 0804.0422}}.

\bibitem{Hirai:2008aj}
{\bf Asymmetry Analysis} Collaboration, M.~Hirai and S.~Kumano,
  ``{Determination of gluon polarization from deep inelastic scattering and
  collider data},'' {\em Nucl. Phys.} {\bf B813} (2009) 106--122,
\href{http://arXiv.org/abs/0808.0413}{{\tt 0808.0413}}.

\bibitem{deFlorian:2009vb}
D.~de~Florian, R.~Sassot, M.~Stratmann, and W.~Vogelsang, ``{Extraction of
  Spin-Dependent Parton Densities and Their Uncertainties},''
\href{http://arXiv.org/abs/0904.3821}{{\tt 0904.3821}}.

\bibitem{Brodsky:1988ip}
S.~J. Brodsky, J.~R. Ellis, and M.~Karliner, ``{Chiral Symmetry and the Spin of
  the Proton},'' {\em Phys. Lett.} {\bf B206} (1988)
309.

\bibitem{Schreiber:1988uw}
A.~W. Schreiber and A.~W. Thomas, ``{SPIN DEPENDENT STRUCTURE FUNCTIONS IN THE
  CLOUDY BAG MODEL},'' {\em Phys. Lett.} {\bf B215} (1988)
141.

\bibitem{Myhrer:1988ap}
F.~Myhrer and A.~W. Thomas, ``{SPIN STRUCTURE FUNCTIONS AND GLUON EXCHANGE},''
  {\em Phys. Rev.} {\bf D38} (1988)
1633.

\bibitem{Wakamatsu:1999nj}
M.~Wakamatsu and T.~Watabe, ``{Spin and orbital angular momentum distribution
  functions of the nucleon},'' {\em Phys. Rev.} {\bf D62} (2000) 054009,
\href{http://arXiv.org/abs/hep-ph/9912500}{{\tt hep-ph/9912500}}.

\bibitem{Bass:2006dq}
S.~D. Bass, ``{Spin constraints on Regge predictions and perturbative evolution
  in high energy collisions},'' {\em Mod. Phys. Lett.} {\bf A22} (2007)
  1005--1012,
\href{http://arXiv.org/abs/hep-ph/0606067}{{\tt hep-ph/0606067}}.

\bibitem{Aharony:1999ti}
O.~Aharony, S.~S. Gubser, J.~M. Maldacena, H.~Ooguri, and Y.~Oz, ``{Large N
  field theories, string theory and gravity},'' {\em Phys. Rept.} {\bf 323}
  (2000) 183--386,
\href{http://arXiv.org/abs/hep-th/9905111}{{\tt hep-th/9905111}}.

\bibitem{Polchinski:2002jw}
J.~Polchinski and M.~J. Strassler, ``{Deep inelastic scattering and
  gauge/string duality},'' {\em JHEP} {\bf 05} (2003) 012,
\href{http://arXiv.org/abs/hep-th/0209211}{{\tt hep-th/0209211}}.

\bibitem{Hatta:2007he}
Y.~Hatta, E.~Iancu, and A.~H. Mueller, ``{Deep inelastic scattering at strong
  coupling from gauge/string duality : the saturation line},'' {\em JHEP} {\bf
  01} (2008) 026,
\href{http://arXiv.org/abs/0710.2148}{{\tt 0710.2148}}.

\bibitem{Hatta:2007cs}
Y.~Hatta, E.~Iancu, and A.~H. Mueller, ``{Deep inelastic scattering off a N=4
  SYM plasma at strong coupling},'' {\em JHEP} {\bf 01} (2008) 063,
\href{http://arXiv.org/abs/0710.5297}{{\tt 0710.5297}}.

\bibitem{BallonBayona:2007qr}
C.~A. Ballon~Bayona, H.~Boschi-Filho, and N.~R.~F. Braga, ``{Deep inelastic
  scattering from gauge string duality in the soft wall model},'' {\em JHEP}
  {\bf 03} (2008) 064,
\href{http://arXiv.org/abs/0711.0221}{{\tt 0711.0221}}.

\bibitem{BallonBayona:2007rs}
C.~A. Ballon~Bayona, H.~Boschi-Filho, and N.~R.~F. Braga, ``{Deep inelastic
  structure functions from supergravity at small x},'' {\em JHEP} {\bf 10}
  (2008) 088,
\href{http://arXiv.org/abs/0712.3530}{{\tt 0712.3530}}.

\bibitem{Cornalba:2008sp}
L.~Cornalba and M.~S. Costa, ``{Saturation in Deep Inelastic Scattering from
  AdS/CFT},'' {\em Phys. Rev.} {\bf D78} (2008) 096010,
\href{http://arXiv.org/abs/0804.1562}{{\tt 0804.1562}}.

\bibitem{Pire:2008zf}
B.~Pire, C.~Roiesnel, L.~Szymanowski, and S.~Wallon, ``{On AdS/QCD
  correspondence and the partonic picture of deep inelastic scattering},'' {\em
  Phys. Lett.} {\bf B670} (2008) 84--90,
\href{http://arXiv.org/abs/0805.4346}{{\tt 0805.4346}}.

\bibitem{Gao:2009ze}
J.-H. Gao and B.-W. Xiao, ``{Polarized Deep Inelastic and Elastic Scattering
  From Gauge/String Duality},''
\href{http://arXiv.org/abs/0904.2870}{{\tt 0904.2870}}.

\bibitem{Wandzura:1977qf}
S.~Wandzura and F.~Wilczek, ``{Sum Rules for Spin Dependent Electroproduction:
  Test of Relativistic Constituent Quarks},'' {\em Phys. Lett.} {\bf B72}
  (1977)
195.

\bibitem{Kodaira:1979pa}
J.~Kodaira, ``{QCD Higher Order Effects in Polarized Electroproduction: Flavor
  Singlet Coefficient Functions},'' {\em Nucl. Phys.} {\bf B165} (1980)
129.

\bibitem{Kodaira:1979ib}
J.~Kodaira, S.~Matsuda, K.~Sasaki, and T.~Uematsu, ``{QCD Higher Order Effects
  in Spin Dependent Deep Inelastic Electroproduction},'' {\em Nucl. Phys.} {\bf
  B159} (1979)
99.

\bibitem{Bjorken:1966jh}
J.~D. Bjorken, ``{Applications of the Chiral U(6) x (6) Algebra of Current
  Densities},'' {\em Phys. Rev.} {\bf 148} (1966)
1467--1478.

\bibitem{Freedman:1998tz}
D.~Z. Freedman, S.~D. Mathur, A.~Matusis, and L.~Rastelli, ``{Correlation
  functions in the CFT($d$)/AdS($d+1$) correspondence},'' {\em Nucl. Phys.}
  {\bf B546} (1999) 96--118,
\href{http://arXiv.org/abs/hep-th/9804058}{{\tt hep-th/9804058}}.

\bibitem{Chalmers:1998xr}
G.~Chalmers, H.~Nastase, K.~Schalm, and R.~Siebelink, ``{R-current correlators
  in N = 4 super Yang-Mills theory from anti-de Sitter supergravity},'' {\em
  Nucl. Phys.} {\bf B540} (1999) 247--270,
\href{http://arXiv.org/abs/hep-th/9805105}{{\tt hep-th/9805105}}.

\bibitem{Babington:2003vm}
J.~Babington, J.~Erdmenger, N.~J. Evans, Z.~Guralnik, and I.~Kirsch, ``{Chiral
  symmetry breaking and pions in non-supersymmetric gauge / gravity duals},''
  {\em Phys. Rev.} {\bf D69} (2004) 066007,
\href{http://arXiv.org/abs/hep-th/0306018}{{\tt hep-th/0306018}}.

\bibitem{Ball:1995ye}
R.~D. Ball, S.~Forte, and G.~Ridolfi, ``{Scale dependence and small x behavior
  of polarized patron distributions},'' {\em Nucl. Phys.} {\bf B444} (1995)
  287--309,
\href{http://arXiv.org/abs/hep-ph/9502340}{{\tt hep-ph/9502340}}.

\bibitem{Bartels:1995iu}
J.~Bartels, B.~I. Ermolaev, and M.~G. Ryskin, ``{Nonsinglet contributions to
  the structure function g1 at small x},'' {\em Z. Phys.} {\bf C70} (1996)
  273--280,
\href{http://arXiv.org/abs/hep-ph/9507271}{{\tt hep-ph/9507271}}.

\bibitem{Bartels:1996wc}
J.~Bartels, B.~I. Ermolaev, and M.~G. Ryskin, ``{Flavor singlet contribution to
  the structure function g1 at small x},'' {\em Z. Phys.} {\bf C72} (1996)
  627--635,
\href{http://arXiv.org/abs/hep-ph/9603204}{{\tt hep-ph/9603204}}.

\bibitem{Heimann:1973hq}
R.~L. Heimann, ``{Spin dependent high frequency inelastic electron scattering
  and helicity flip couplings},'' {\em Nucl. Phys.} {\bf B64} (1973)
429--463.

\bibitem{Bianchi:2001cm}
M.~Bianchi, S.~Kovacs, G.~Rossi, and Y.~S. Stanev, ``{Properties of the Konishi
  multiplet in N = 4 SYM theory},'' {\em JHEP} {\bf 05} (2001) 042,
\href{http://arXiv.org/abs/hep-th/0104016}{{\tt hep-th/0104016}}.

\bibitem{Jaffe:1989jz}
R.~L. Jaffe and A.~Manohar, ``{The G(1) Problem: Fact and Fantasy on the Spin
  of the Proton},'' {\em Nucl. Phys.} {\bf B337} (1990)
509--546.

\bibitem{Kotikov:2002ab}
A.~V. Kotikov and L.~N. Lipatov, ``{DGLAP and BFKL evolution equations in the
  N=4 supersymmetric gauge theory},'' {\em Nucl. Phys.} {\bf B661} (2003)
  19--61,
\href{http://arXiv.org/abs/hep-ph/0208220}{{\tt hep-ph/0208220}}.

\bibitem{Brower:2006ea}
R.~C. Brower, J.~Polchinski, M.~J. Strassler, and C.-I. Tan, ``{The Pomeron and
  Gauge/String Duality},'' {\em JHEP} {\bf 12} (2007) 005,
\href{http://arXiv.org/abs/hep-th/0603115}{{\tt hep-th/0603115}}.

\bibitem{Brower:2007xg}
R.~C. Brower, M.~J. Strassler, and C.-I. Tan, ``{On The Pomeron at Large 't
  Hooft Coupling},'' {\em JHEP} {\bf 03} (2009) 092,
\href{http://arXiv.org/abs/0710.4378}{{\tt 0710.4378}}.

\bibitem{Brower:2008cy}
R.~C. Brower, M.~Djuric, and C.-I. Tan, ``{The Kalb-Ramond Odderon in
  AdS/CFT},''
\href{http://arXiv.org/abs/0812.0354}{{\tt 0812.0354}}.

\bibitem{Kotikov:2004er}
A.~V. Kotikov, L.~N. Lipatov, A.~I. Onishchenko, and V.~N. Velizhanin,
  ``{Three-loop universal anomalous dimension of the Wilson operators in N = 4
  SUSY Yang-Mills model},'' {\em Phys. Lett.} {\bf B595} (2004) 521--529,
\href{http://arXiv.org/abs/hep-th/0404092}{{\tt hep-th/0404092}}.

\bibitem{D'Hoker:1998gd}
E.~D'Hoker and D.~Z. Freedman, ``{Gauge boson exchange in AdS(d+1)},'' {\em
  Nucl. Phys.} {\bf B544} (1999) 612--632,
\href{http://arXiv.org/abs/hep-th/9809179}{{\tt hep-th/9809179}}.

\bibitem{Ji:1996ek}
X.-D. Ji, ``{Gauge invariant decomposition of nucleon spin},'' {\em Phys. Rev.
  Lett.} {\bf 78} (1997) 610--613,
\href{http://arXiv.org/abs/hep-ph/9603249}{{\tt hep-ph/9603249}}.

\bibitem{Anselmi:1998ms}
D.~Anselmi, ``{The N = 4 quantum conformal algebra},'' {\em Nucl. Phys.} {\bf
  B541} (1999) 369--385,
\href{http://arXiv.org/abs/hep-th/9809192}{{\tt hep-th/9809192}}.

\bibitem{Abe:1997cx}
{\bf E154} Collaboration, K.~Abe {\em et al.}, ``{Precision determination of
  the neutron spin structure function g1(n)},'' {\em Phys. Rev. Lett.} {\bf 79}
  (1997) 26--30,
\href{http://arXiv.org/abs/hep-ex/9705012}{{\tt hep-ex/9705012}}.

\end{thebibliography}\endgroup

\end{document}